\documentclass{emulateapj}
\usepackage{color}
\usepackage{latexsym, graphicx, amssymb, longtable, epsf}
\long\def\symbolfootnote[#1]#2{\begingroup\def\thefootnote{\fnsymbol{footnote}}
\footnote[#1]{#2}\endgroup}

\shorttitle{Asymmetric Orbital Distribution near Mean Motion Resonance}
\shortauthors{Xie}

\begin{document}
%\begin{CJK*}{UTF8}{gbsn}

\title{Asymmetric Orbital Distribution near Mean Motion Resonance: Application to Planets Observed by \emph{Kepler} and Radial Velocities}

%\author{Ji-Wei Xie (谢基伟)$^{1, 2}$}
\author{Ji-Wei Xie $^{1, 2}$}
\affil{$^1$Department of Astronomy \& Key Laboratory of Modern Astronomy and Astrophysics in Ministry of Education, Nanjing University, Nanjing, 210093, China; jwxie@nju.edu.cn}
\affil{$^2$Department of Astronomy and Astrophysics, University of Toronto, Toronto, ON M5S 3H4, Canada; jwxie@astro.utoronto.ca}

%%%%%%%%%%%%%%%%%%%%%%%%%%%%%%%%%%%%%%%%%%%%%%%%%%%%%%%%%%%%%%%%%%%

\begin{abstract}
Many multiple planet systems have been found by the \emph{Kepler} transit survey and various Radial Velocity (RV) surveys. \emph{Kepler} planets show an asymmetric feature, namely there are small but significant deficits/excesses of planet pairs with orbital period spacing slightly narrow/wide of the exact resonance, particularly near the first order Mean Motion Resonance (MMR), such as 2:1 and 3:2 MMR. Similarly, if not exactly the same, an asymmetric feature (pileup wide of 2:1 MMR) is also seen in RV planets, but only for massive ones.  

%We explore in this paper that how these observations could be understood in a common context.

%
We analytically and numerically study planets' orbital evolutions near/in MMR. We find that their orbital period ratios could be asymmetrically distributed around the MMR center regardless of dissipation. In the case of no dissipation, \emph{Kepler} planets' asymmetric orbital distribution could be partly reproduced for 3:2 MMR but not for 2:1 MMR, implying dissipation might be more important to the latter. The pileup of massive RV planets just wide of 2:1 MMR is found to be consistent with the scenario that planets formed separately then migrated toward MMR. The location of the pileup infers a $K$ value of 1-100 on order of magnitude for massive planets, where $K$ is the damping rate ratio between orbital eccentricity and semimajor axis during planet migration.  
\end{abstract}

%%%%%%%%%%%%%%%%%%%%%%%%%%%%%%%%%%%%%%%%%%%%%%%%%%%%%%%%%%%%%%%%%%%

\section{Introduction}
The \emph{Kepler} mission has discovered from its first 16 months data  over $2300$ planetary candidates \citep{Bor11, Bat12}.  Over one third ($>$800) of these candidates are in multiple transiting candidate planetary systems, and one remarkable feature of them, as shown by  \citet{Lis11} and \citet{Fab12a}, is that the vast majority of candidate pairs are neither in nor near low-order mean motion resonance (MMR hereafter, see also in \citet{VF12}), however there are small but significant excesses/deficits of candidate pairs slightly wider/narrow of the exact resonance (or nominal resonance center), particularly near the first order MMR, such as 2:1 and 3:2 MMR.  

Such an intriguing asymmetric period ratio distribution has stimulated a number of theorists recently, who developed different models to understand and interpret it. \citet{LW12, BM12, Del12} consider that such an asymmetric period ratio distribution around MMR could be an outcome of resonant couples having underwent eccentricity damping during some dissipative evolutions, such as tidal dissipation (see also in \citet{TP07}). On the other side, \citet{Rei12} attempts to  interpret it as a result of the combination of stochastic and smooth planet migrations.  

Beside and before the \emph{Kepler} transit survey, many near MMR planets had been found by various Radial Velocity (RV hereafter) surveys. As we will show below (section \ref{dis_rv}), similar, if not exactly the same, features of the period ratio distributions seen in \emph{Kepler} planets,  have been also shown in RV planets. One question is how all these features/clues in both the \emph{Kepler} and RV samples could be understood systematically in a common context. This paper is such an attempt and it is organized as the following.

We first analytically study the dynamics of planets near/in MMR in section \ref{sec_analytic}, and confirm the analytical results with numerical simulations in section \ref{sec_numerical}. We find that planets' orbital distribution could be asymmetric around the MMR center under certain conditions. We then discuss its implications to \emph{Kepler} and RV planets in section \ref{dis}.  Finally, we summarize this paper in section \ref{sum}. Some analytical derivations are also given in the appendix A and B as supplementary.  We note that \cite{Pet12} posted their paper to arxiv.org just a few days before submitting this paper, which, independently and in a different way, arrived at many of the results presented in this paper.

\section{Asymmetric orbit distribution near MMR}
We study the orbital evolutions of two planets (orbiting a central star) near/in first order MMR. As we will show below, the orbit distribution could be asymmetric near the MMR center under certain circumstances. 
\subsection{Analytic Study} 
\label{sec_analytic}
\subsubsection{No dissipation (analytical)}
\label{nod1}
For simplicity, we assume both planets' orbits are coplanar. The total energy, or Hamiltonian, is \citep{MD99} 
\begin{equation}
H = -\frac{GM_{\star}m_{1}}{2a_{1}} -\frac{GM_{\star}m_{2}}{2a_{2}}-\frac{Gm_{1}m_{2}}{a_{2}}R^{j},
\end{equation}
where $G$ is the gravity constant, $M_{\star}$ is the stellar mass, and following \citet{Lit12}, the disturbing function due to the $j:j-1$ resonance is 
\begin{equation}
R^{j} = f_{1}e_{1}cos(\phi_{1})+f_{2}e_{2}cos(\phi_{2}),
\label{rj}
\end{equation}
where
\begin{equation}
\phi_{1} = \lambda^{j}-\varpi_{1}, \, \phi_{2} = \lambda^{j}-\varpi_{2},
\end{equation}
are the two resonance angles for
\begin{equation}
\lambda^{j} = j\lambda_{2}-(j-1)\lambda_{1}.
\end{equation}
Hereafter, we adopt the convention that properties with subscripts ``1'' and ``2'' belong to the inner and outer planets respectively. In the above, \{$m$, $a$, $e$, $\lambda$, $\varpi$\} are the mass and standard orbital elements for planets. $f_{1}$ and $f_{2}$ are relevant Laplace coefficients, which are on order of unity and  tabulated in \citet{MD99} and \citet{Lit12}.

Using the Lagrange's planetary equation (on the lowest order terms in $e$), we derive the evolutions of planets' semi major axes and eccentricities,
\begin{eqnarray}
\dot{a_{1}}&=&-2(j-1)\frac{Gm_{2}}{n_{1}a_{1}a_{2}}(f_{1}e_{1}sin\phi_{1}+f_{2}e_{2}sin\phi_{2}), \nonumber \\
\dot{a_{2}}&=&2j\frac{Gm_{1}}{n_{2}a_{2}^{2}}(f_{1}e_{1}sin\phi_{1}+f_{2}e_{2}sin\phi_{2}),
\label{adot0}
\end{eqnarray}
\begin{eqnarray}
\dot{e_{1}} &=& \frac{Gm_{2}}{n_{1}a_{1}^{2}a_{2}e_{1}}(f_{1}e_{1}sin\phi_{1}), \nonumber \\
\dot{e_{2}} &=& \frac{Gm_{1}}{n_{2}a_{2}^{3}e_{2}}(f_{2}e_{2}sin\phi_{2}),
\label{edot}
\end{eqnarray}
where, $n_{1}$ and $n_{2}$ are the mean motion of the inner and outer planets respectively.

Using equation \ref{edot} to eliminate $\phi_{1}$ and $\phi_{2}$, we can rewrite equation \ref{adot0} as 
\begin{eqnarray}
\frac{\dot{a_{1}}}{a_{1}}&=&-2(j-1)\left(e_{1}\dot{e_{1}}+q\rho^{\frac{1}{3}}e_{2}\dot{e_{2}}\right) \nonumber \\
\frac{\dot{a_{2}}}{a_{2}}&=&2j\left(q^{-1}\rho^{-\frac{1}{3}}e_{1}\dot{e_{1}}+e_{2}\dot{e_{2}}\right)
\label{adot1}
\end{eqnarray}
which integrate to give
\begin{eqnarray}
{\rm ln}\, a_{1}+(j-1)\left(e_{1}^{2} +q\rho^{\frac{1}{3}}e_{2}^{2}\right) = {\rm Const.}\nonumber \\
{\rm ln}\,a_{2}-j\left(q^{-1}\rho^{-\frac{1}{3}}e_{1}^{2}+e_{2}^{2}\right) = {\rm Const.}
\label{const1}
\end{eqnarray}
where we have defined
\begin{eqnarray}
\rho = j/(j-1) \, \, \, {\rm and} \, \, \, q=m_{2}/m_{1}
\end{eqnarray}
We note equations \ref{adot1} or \ref{const1} are equivalent to the well known constants of motion in resonance (see appendix A).
A worth noting implication of equation \ref{adot1} or \ref{const1} is that if planet pairs initially formed with circular orbit near MMR, they will shift to a little bit larger orbital period ratio as their eccentricities are excited, inducing an asymmetric orbit distribution near MMR (see numerical confirmation in section \ref{sec_numerical}). Using equation \ref{adot1}, this small shift extent in period ratio ($p_{2}/p_{1}$) can be estimated as
\begin{eqnarray}
\label{dp1}
{\rm d}\left(\frac{p_{2}}{p_{1}}\right) &=&\frac{3}{2}\left(\frac{a_{2}}{a_{1}}\right)^{3/2}\left(\frac{{\rm d}a_{2}}{a_{2}}-\frac{{\rm d}a_{1}}{a_{1}}\right) \\
& = & \frac{3}{2}j\left[\left(q^{-1}\rho^{\frac{2}{3}}+1\right){\rm d}e_{1}^{2} + \rho\left(q\rho^{-\frac{2}{3}}+1\right){\rm d}e_{2}^{2}\right] \nonumber
\end{eqnarray}
According to \citet{MD99} (see their Eqn. 8.209 and 8.210), the maximum eccentricity increase in $e_{1}$ and $e_{2}$ (or critical eccentricities) are 
\begin{eqnarray}
e_{\rm cr_1} &=& \sqrt{6}\left|\frac{3}{f_{1}}(j-1)^{\frac{4}{3}}j^{\frac{2}{3}}+j^{\frac{4}{3}}(j-1)^{\frac{2}{3}}/q\right|^{-1/3}\left(\frac{m_{2}}{M_{\star}}\right)^{1/3} \nonumber \\
e_{\rm cr_2} &=& \sqrt{6}\left|\frac{3}{f_{2}}(j-1)^{\frac{4}{3}}j^{\frac{2}{3}}q+j^{2}\right|^{-1/3}\left(\frac{m_{1}}{M_{\star}}\right)^{1/3}.
\label{ecr}
\end{eqnarray}
Setting ${\rm d}e_{1}^{2}=e_{\rm cr_1}^{2}$ and ${\rm d}e_{2}^{2}=e_{\rm cr_2}^{2}$, then equation \ref{dp1} will give an estimate of the largest asymmetric shift of period ratio.

%%%%%%%%%%
\subsubsection{With dissipation (analytical)}
\label{dissipation1}
Dissipation processes (e.g., tidal evolution, disk migration) may play an import role during planet formation and evolution. Generally they cause changing on planets' orbital semi major axes and damping in eccentricities. To include these effects, we consider the following changing/damping terms (i.e., inverse of the damping timescales) of semi major axes and eccentricities,
\begin{eqnarray}
\gamma_{\rm ak}=-\frac{1}{a_{\rm k}}\frac{{\rm d}a_{\rm k}}{{\rm d}t}, \, \, \gamma_{\rm ek}=-\frac{1}{e_{\rm k}}\frac{{\rm d}e_{\rm k}}{{\rm d}t},
\end{eqnarray}
where (hereafter) $\rm k=1,2$ for the inner and outer planets respectively. Note, $\gamma_{\rm ak}$ could be negative, which indicates outward migration, and $\gamma_{\rm ek}$ is generally positive, i.e., eccentricity is damped in dissipation process.   

Following \citet{Lit12} (see the appendix B for the derivation), the evolutions of the semi major axes of two planets (after adding above damping terms) are.
\begin{eqnarray}
\frac{\dot{a}_{\rm 1}}{a_{\rm 1}}&=& -\frac{2}{jq\rho^{2/3}}\frac{1}{\Delta^{2}}\left(\frac{m_{2}}{M_{\star}}\right)^{2}\left(q\rho^{\frac{1}{3}}f_{1}^{2}\gamma_{\rm e1}+f_{2}^{2}\gamma_{\rm e2}\right) - \gamma_{\rm a1},
\nonumber \\
\frac{\dot{a}_{\rm 2}}{a_{\rm 2}} &=& \frac{2}{j}\frac{1}{\Delta^{2}}\left(\frac{m_{1}}{M_{\star}}\right)^{2}\left(q\rho^{\frac{1}{3}}f_{1}^{2}\gamma_{\rm e1}+f_{2}^{2}\gamma_{\rm e2}\right) - \gamma_{\rm a2},
\label{da4}
\end{eqnarray}
where 
\begin{eqnarray}
\Delta = \frac{j-1}{j}\frac{p_{2}}{p_{1}}-1
\end{eqnarray}
is the proximity to the nominal resonance center, and thus its evolution follows,
\begin{eqnarray}
\dot{\Delta}& = &\frac{3}{2}\left(\frac{\dot{a}_{\rm 2}}{a_{\rm 2}}-\frac{\dot{a}_{\rm 1}}{a_{\rm1}}\right)
= \frac{3}{j\Delta^{2}}\left(\frac{m_{1}}{M_{\star}}\right)^{2}\left(1+q\rho^{-\frac{2}{3}}\right)\times \nonumber \\
&&\left(q\rho^{\frac{1}{3}}f_{1}^{2}\gamma_{\rm e1}+f_{2}^{2}\gamma_{\rm e2}\right)+\gamma_{a1}-\gamma_{a2}
\end{eqnarray}

If $\gamma_{\rm a1}\ge\gamma_{\rm a2}$, then $\dot{\Delta}$ will be always positive, namely the two planets will always keep divergent migration, i.e., their period ratio will always increase. This is the case if the planetary system undergoes 
 tidal evolution \citep{TP07, LW12, BM12}.
 
If $\gamma_{\rm a1}<\gamma_{\rm a2}$ otherwise, then there is an stable equilibrium with 
\begin{eqnarray}
\Delta_{\rm eq}= \frac{m_{1}}{M_{\star}}\left[\frac{3\left(1+q\rho^{-\frac{2}{3}}\right)\left(q\rho^{\frac{1}{3}}f_{1}^{2}\gamma_{\rm e1}+f_{2}^{2}\gamma_{\rm e2}\right)}{j\left(\gamma_{a2}-\gamma_{a1}\right)}\right]^{1/2},
\label{deq}
 \end{eqnarray}
 wider/narrower than which, the two planets will undergo convergent/divergent migration, thus eventually they will be locked at $\Delta=\Delta_{\rm eq}$.
 
Interestingly, the above equation can be roughly written as 
 \begin{eqnarray}
 \Delta_{\rm eq}\sim \mu K^{1/2},
 \label{deq2}
 \end{eqnarray}
 where $\mu$ is the typical planet-star mass ratio of the system and $K=\gamma_{\rm e}/\gamma_{\rm a}$ is the well known model parameter describing the ratio between the damping rate of orbital eccentricity and that of semimajor axis. From equation (\ref{deq2}), we see that the theoretical parameter $K$ is linked to an observable $\Delta_{\rm eq}$. We will discuss this more in section \ref{dis_rv}.

%%%%%%%%%%%%%%%%%%%%%%
\subsection{Numerical Study} 
\label{sec_numerical}
For comparison against the above analytical results, we perform some 3-Body (1 star + 2 planets) simulations using the well-tested N-body integrator MERCURY \citep{CM97}. For all the simulations, the central star is set with a mass $M_{\star}=M_{\odot}$, and all angular orbital elements, except for orbital inclinations, are initially randomly set. For most simulations, the semi major axis of the inner planet is set at 0.1 AU if not specified.
\subsubsection{No dissipation (numerical)}
\label{nod2}
From equations \ref{adot1}-\ref{dp1}, we expect that planets' orbits have an asymmetric distribution near the MMR center. Here, we numerically show such an asymmetry and its dependence on the initial period ratio, orbital eccentricities, inclinations and planetary masses.

\begin{figure*}
\begin{center}
\includegraphics[width=0.9\textwidth]{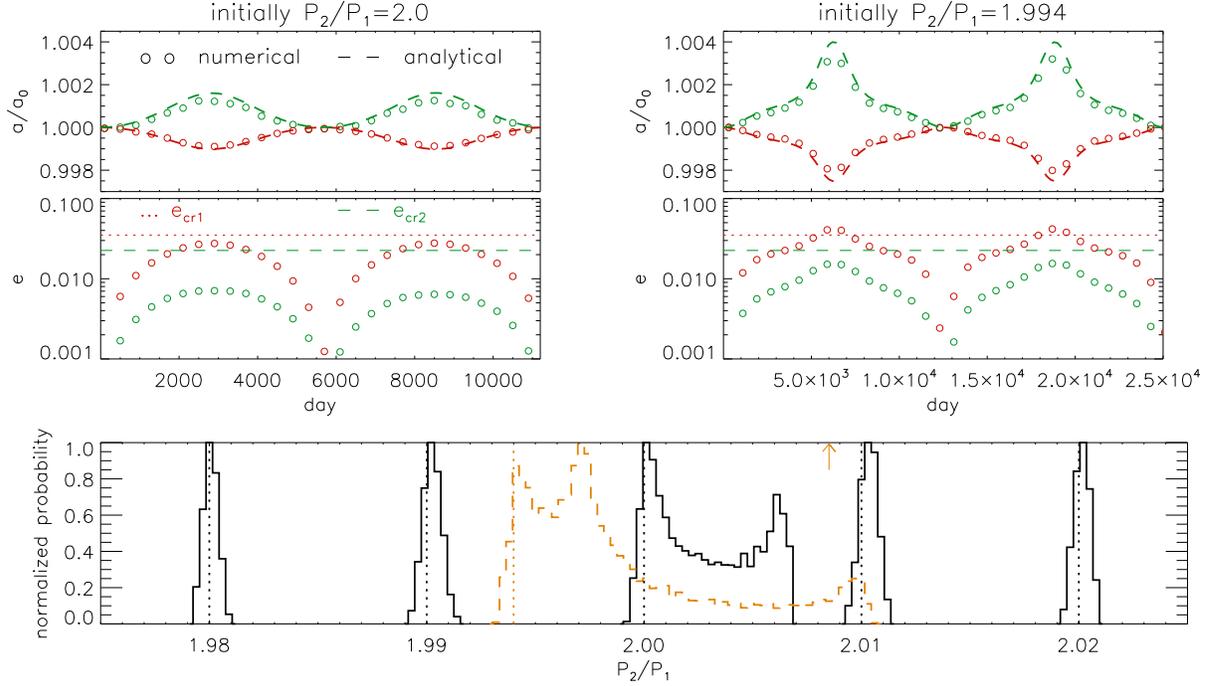}
  \caption{Evolutions of semimajor axes (top, normalized to the initial value, $a/a_{\rm 0}$),  and eccentricity (middle, $e$) of two planets with masses $m_{1}=m_{2}=10M_{\oplus}$ (red for the inner planet and green for the outer one) initial orbital ratio of $p_{\rm 2}/p_{\rm1}=2.0$ (top left 2 panels) and $p_{\rm 2}/p_{\rm1}=1.994$ (top right 2 panels), circular $e_{\rm 1}=e_{\rm 2}=0$) and coplanar ($i_{\rm 12}=0$) orbits.  In the top 2 panels, the circles are numerical results and the dashed lines are analytical results based on equation \ref{const1}. In the bottom two panels, the horizontal dot lines mark the critical eccentricities (Eqn.\ref{ecr}). 
Performing above simulation 100 times with random initial angular orbital elements, we plot the average orbital ratio (sampled at uniformly-spaced time points) distributions in the bottom panel for the cases with initial $p_{\rm 2}/p_{\rm1}=1.98, 1.99, 1.994, 2.0 , 2.01 {\rm \, and \, } 2.02$.  A dot vertical line is plot in each histogram to indicate the initial period ratio. The asymmetric orbital ratio distribution is most prominent at a little bit narrower than the MMR center (i.e., $p_{\rm 2}/p_{\rm1}=1.994$ here) and become weaker and weaker as it is further away from MMR. The arrow in the bottom panel marks the largest orbital ratio shift estimated from equations \ref{dp1}-\ref{ecr}, which is roughly consistent with the numerical results (orange histogram). }
\label{fig_asym}
   \end{center}
\end{figure*}

Figure \ref{fig_asym} shows the orbital evolutions of two equal mass ($10M_{\oplus}$) planets initially with circular and coplanar orbits but different orbital ratios. Planets' semimajor axes and eccentricities follow periodical oscillations, and their period ratios increase with eccentricities as expected from equation \ref{adot1}. On average the planets spend more time on orbits wider than the initial ones, causing an asymmetric distribution in their period ratio.  The asymmetry become weaker as the planet pair is further away from MMR. However, the most prominent asymmetric feature is not realized at the MMR center but at a little bit narrower than the center. The reason is that planets' eccentricities get most excited when they are at the separatrix which is at narrower than the nominal resonance center for the first order MMR \citep{MD99}. 

\begin{figure*}
\begin{center}
\includegraphics[width=\textwidth]{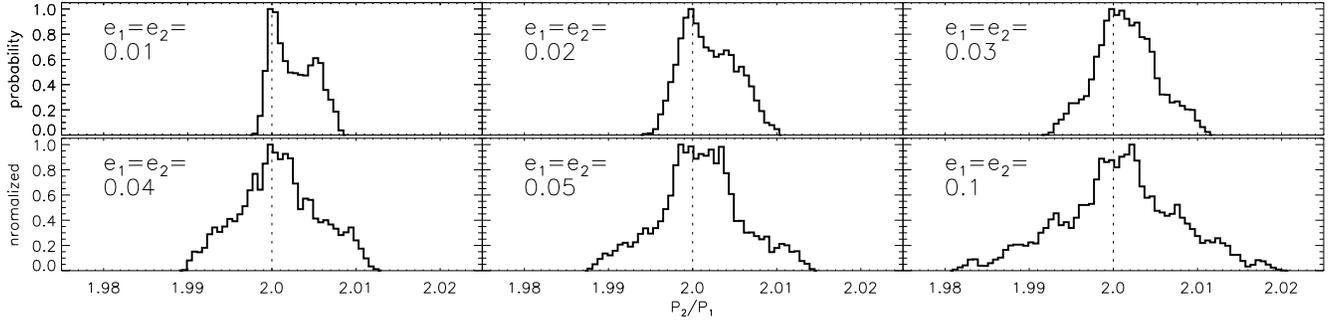}
  \caption{Similar to the bottom panel of figure \ref{fig_asym} but here we compare the period ratio distributions of cases with different initial eccentricities (printed in each panel). As shown, the asymmetric feature diminishes as the eccentricity become comparable to or larger than the critical eccentricities. Here, $e_{\rm cr_{1}} = 0.035, e_{\rm cr_{2}} = 0.023$ according to equations \ref{ecr}.   }
\label{fig_ecc}
   \end{center}
\end{figure*}

Figure \ref{fig_ecc} shows how the asymmetry is affected by the initial orbital eccentricities. As expected from equation \ref{adot1}, if the eccentricity is initially larger, then it will have larger possibility (compared to the case of zero initial eccentricity) to decrease in the future, thus the period ratio will become more symmetric around the initial one.  The critical eccentricity, greater than which the asymmetry will be very weak, could be estimated using equation \ref{ecr}, which is consistent with the numerical results and the results within the context of the restricted 3-body problem \citep{MD99}.

Figure \ref{fig_mas} shows the effect of planetary mass on the asymmetry. As expected from equation \ref{adot1}-\ref{ecr}, increasing mass leading to larger eccentricity excitation and thus larger period ratio shift extent. Roughly, systems with similar total masses (regardless of mass ratio) have similar shift extents. 

\begin{figure*}
\begin{center}
\includegraphics[width=\textwidth]{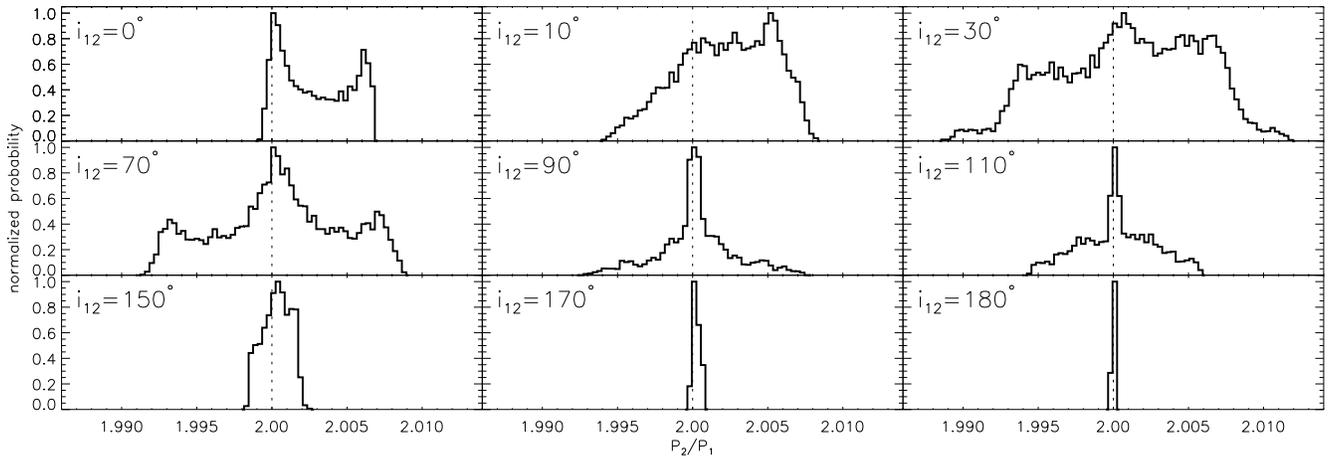}
  \caption{Similar to Fig.\ref{fig_ecc}, but here we investigate the dependance of asymmetry on the initial relative orbital inclination ($i_{\rm 12}$) of the two planets. As shown, the asymmetry become very weak if $i_{\rm 12}>10^{\circ}$. }
\label{fig_inc}
   \end{center}
\end{figure*}

Figure \ref{fig_inc} shows the role of relative inclination $i_{12}$ in the asymmetry. Generally the asymmetry becomes very weak for $i_{12}>10^{\circ}$. This is not surprised, as the above analytical studies are all based on an assumption of low $i_{12}$. For large $i_{12}$, more terms ( e.g., on the oder of $ie$) should be considered in the disturbing function in equation \ref{rj}, and in such cases, planets could be involved in second order of MMR, which is symmetric around nominal resonance center \citep{MD99}. %

\subsubsection{With dissipation (numerical)}
\label{dissipation2}
The case of divergent migration (i.e., $\gamma_{a1}\ge\gamma_{a2}$) has been studied recently in detail recently by \citet{TP07, LW12, BM12}. Here we focus on the other case where $\gamma_{a1}<\gamma_{a2}$. For simplicity, we assume $\gamma_{a1}=\gamma_{e1}=0$, and damping is only added on the outer planet with $\gamma_{a2}=10^{-8} \, \rm d^{-1}$ and $\gamma_{e2}=K\gamma_{a2}$. The two planets are started at 0.2 and 0.35 AU respectively with an initial orbital period $\sim 2.3$.  We study 7 different K values from 0 to 10000 and 3 different planetary mass sets. The results are plotted in figure \ref{fig_mig}. 

The left 4 panels of figure \ref{fig_mig} plot the results of one simulation with $m_{1}=m_{2}=100M_{\oplus}$ and $K=100$. The out planet moves inward and captures into 2:1 MMR with the inner planet at about $t=2\times10^{7}$ d. After that,  the two planets still moving inward together but with resonance angles, eccentricities and period ratios reaching a relatively stable state. The orbital period ratio at the later state is asymmetric around the nominal MMR center, and the its mean value is roughly consistent with the analytical estimate from equation \ref{deq}. 

The  right panel of figure \ref{fig_mig} shows how $\Delta_{\rm eq}$ depends on planetary mass and damping ratio $K$. Generally we see that $\Delta_{\rm eq}$ is proportional to planetary mass and increases with $K$. Not surprised, the analytical predictions are consistent with the numerical simulations only for relative large $\Delta_{\rm eq}$ and $K$ (see Appendix B). For low $K$ values ($K<10$), $\Delta_{\rm eq}$ do not approach zero but become a positive constant which is proportional to planetary mass. Such a tiny constant $\Delta_{\rm eq}$ may reflect the intrinsic asymmetry of the MMR. However, we note that here the constant $\Delta_{\rm eq}$ is much smaller than the maximum asymmetry estimated by equations \ref{dp1} and \ref{ecr}, which is reasonable because large eccentricity leads to weak asymmetry as seen in figure \ref{fig_ecc}.
 
 %%%%%%%%%%%%%%%%%%%%%%%%%%%%%%%%%%%%
\section{Discussions}
\label{dis}
\subsection{Application to \emph{Kepler} Planets}
\label{dis_kep}
The period ratio distribution of \emph{Kepler} multiple planet candidate systems show an intriguing asymmetric feature near MMR, especially for 2:1 and 3:2 MMR, namely there are small deficits/excesses just  a little bit narrow/wide of the nominal MMR center \citep{Lis11, Fab12a}.   To interpret such an asymmetric feature,  \citet{LW12, BM12} consider that it could be a result of planets undergoing some dissipative evolution, such as tidal dissipation. In such a case, as discussed in section \ref{dissipation1},  $\gamma_{a1}>\gamma_{a2}$, thus the planet period will always increase. To quantitively explain the observed asymmetric period ratio distribution, one needs to put a right amount of dissipation on them. In addition, as tidal effect is only efficient for short period planet, e.g., less 10 days, one needs to resort to other dissipations at larger orbital period where the observed asymmetry is still significant. \citet{Rei12} then considers if the observed period ratio is consistent with the scenario of planets migrating in disks. First, he considers smooth migration and finds that the excess or pileup of planet pairs is too large and too close to the MMR center. His result is expected from our analytical results in figure \ref{fig_mig} and equation \ref{deq}, which shows $\Delta_{\rm eq} \sim 10^{-4}$ (2 order of magnitude lower than the observed one) if assuming a typical \emph{Kepler} planet mass on order of 10 $M_{\oplus}$ and $K=10$. Nevertheless, he further shows that by including certain amount stochastic forces due to disk turbulence during migration, the large pileup at MMR center could be smeared out and a period ratio distribution similar to that of \emph{Kepler} planets could be reproduced.

\begin{figure*}
\begin{center}
\includegraphics[width=0.45\textwidth]{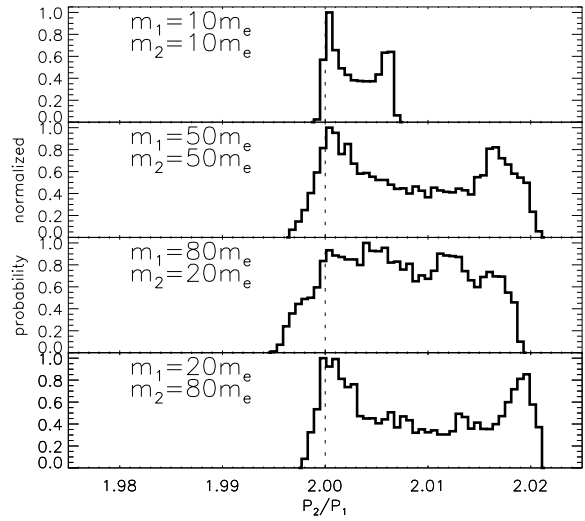}
  \caption{Similar to Fig.\ref{fig_ecc}, but here we investigate the dependance of asymmetry on planets' masses.  As expected from equations \ref{dp1}-\ref{ecr}, the period ratio shift extent increases with planetary mass, and systems with the same total mass (regardless of the mass ratio) have a similar period ratio shift extent.  }
\label{fig_mas}
   \end{center}
\end{figure*}

All the above attempts belong to the case with dissipation. As we have shown (section \ref{nod1} and \ref{nod2}), the period ratio distribution is intrinsically asymmetric near the MMR center even if there is no dissipation. In order to see whether and how the intrinsic asymmetry can reproduce \emph{Kepler} planets' period ratio distribution, we perform the following N-body simulations. Specifically, we draw 4000 planets pairs initially with a uniform period ratio distribution near MMR,  Rayleigh eccentricity and inclination distributions, and uniformly random distribution for all the other angular orbital elements. We use the MERCURY integrator to simulate these 4000 systems individually on a timescale of $10^{5}$ days and intensively output their period ratio very 200 days. The final period ratio distribution is calculated with these output period ratios of all 4000 systems. As \emph{Kepler} multiple planet systems are believed to be highly coplanar within a few degree \citep{Fab12a}, we assume the mean inclination $<i>=2.5^{\circ}$. For simplicity, we only study equal mass pairs, i.e, $m_{1}=m_{2}$ because different mass ratios lead to similar results as long as their total masses are the same (Fig.\ref{fig_mas}).  

\begin{figure*}
\begin{center}
\includegraphics[width=\textwidth,angle=0]{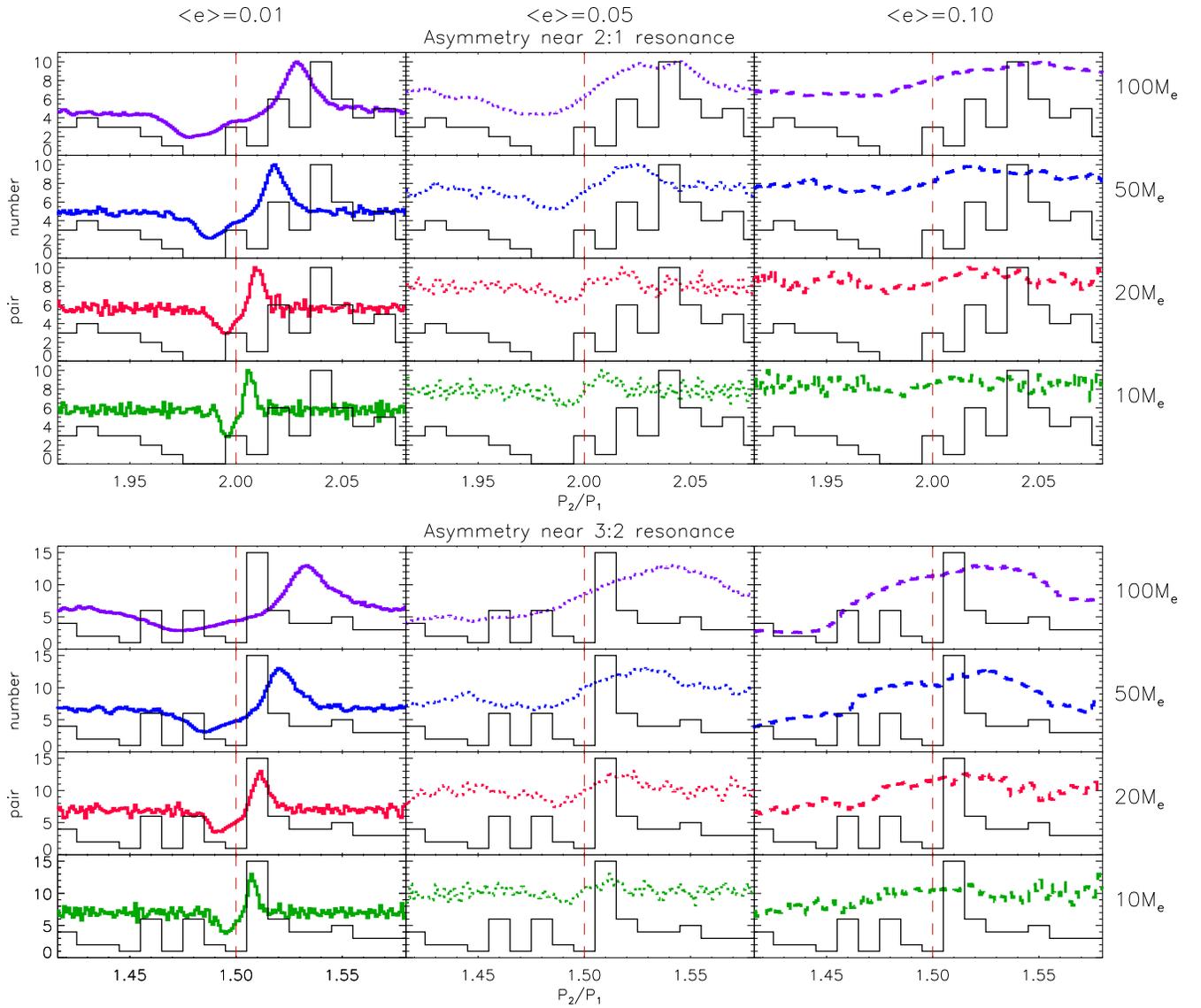}
  \caption{Orbital period ratio ($p_{\rm 2}/p_{\rm1}$) distributions: comparison between simulations (colourized curves, normalized to the same peak as the observational histogram) to \emph{Kepler} observations for planets near 2:1 resonance (top four rows) and those near 3:2 resonance (bottom four rows). For each panel above, we numerically integrate the orbital evolutions of a sample of 4000 planet pairs with an uniform distribution of initial $p_{\rm 2}/p_{\rm1}$ around the nominal resonance center, with equal mass (from bottom to top: 10 $M_{\oplus}$-green, 20 $M_{\oplus}$-red, 50 $M_{\oplus}$-blue, and 100 $M_{\oplus}$-purple), with a Rayleigh distribution of initial orbital eccentricity (from left to right: $<e>=0.01$-solid, $<e>=0.05$-dot, and $<e>=0.1$-dashed).}
\label{fig_obs}
   \end{center}
\end{figure*}

Figure \ref{fig_obs} compares the observed period ratio distribution to those from above simulations with different planetary masses from $10M_{\oplus}$ to $100 M_{\oplus}$ and mean eccentricities from $<e>=0.01$ to $<e>=0.1$. The simulated period ratio distributions have an asymmetric feature resembling the observation, i.e., a trough/pile up just  a little bit narrow/wide of MMR center.  As expected (Fig.\ref{fig_ecc} and \ref{fig_mas}), the asymmetric feature become weaker with increasing eccentricity and more extended with increasing mass. In order to reproduce the observed period ratio distribution, it requires a mean eccentricity less than a few percents and planetary mass about 10-20 $M_{\oplus}$ for 3:2 MMR and $\sim 100 M_{\oplus}$ for 2:1 MMR. The eccentricity requirement is consistent with recent eccentricity estimate with transit timing variation \citep{Fab12b, WL12}. As for the typical mass of \emph{Kepler} planets, it is expected to be 4-9 $M_{\oplus}$ given the typical radii of 2-3 $R_{\oplus}$ and a mass radio distribution either based on fitting of the solar system, $m=M_{\oplus}(r_{\rm}/R_{\oplus})^{2.06}$ \citep{Lis11}, or transit timing variation, $m=3M_{\oplus}(r_{\rm }/R_{\oplus})$ \citep{WL12}. Even considering a relatively large uncertain of mass measurements, say 100\%, such an expected mass is still too low to meet the requirement for 2:1 MMR, although it is comparable to the mass requirement for 3:2 MMR. Therefore, we conclude that the intrinsic MMR asymmetry (without any damping) could partially explain \emph{Kepler} planets' asymmetric period ratio distribution near 3:2 MMR but not 2:1 MMR. For the latter, other mechanisms, e.g., dissipation, should play a more important role.

\subsection{Application to \emph{RV} Planets}
\label{dis_rv}
\begin{figure*}
\begin{center}
\includegraphics[width=0.99\textwidth,angle=0]{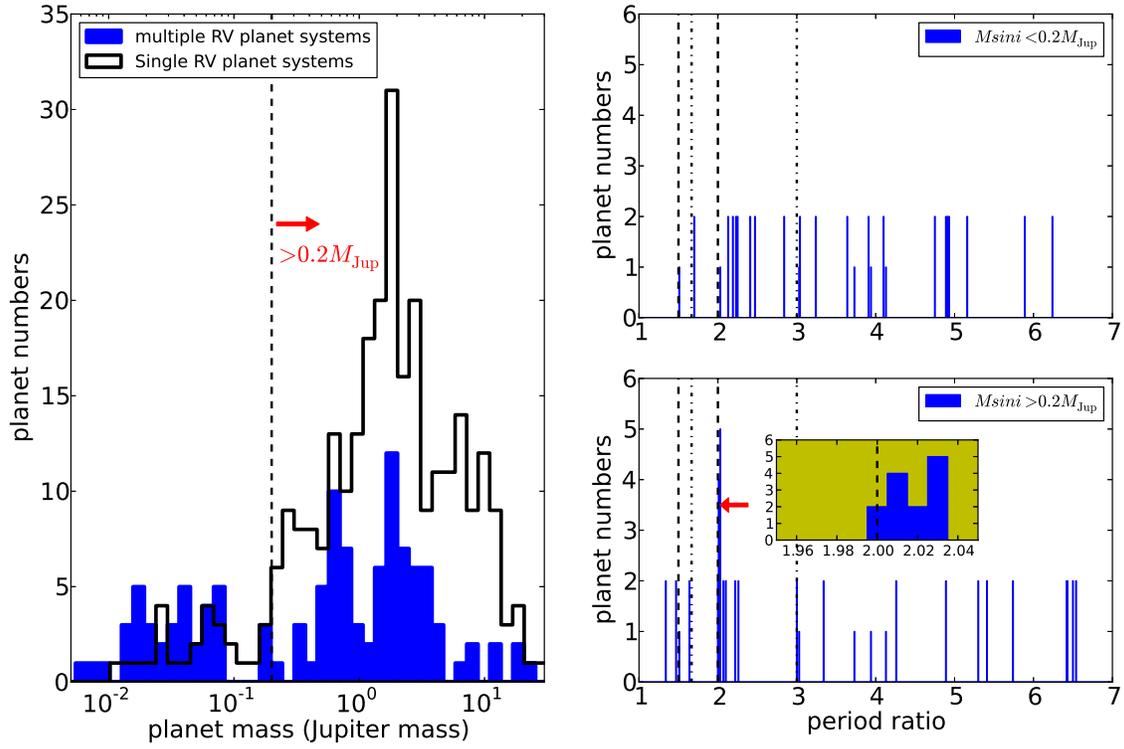}
  \caption{Mass distributions (left panel) and period ratio distributions (right two panels) of  RV planet sample based on the current exoplanet data set from ``exoplanet.org''.  As can be seen, the mass distribution seem bimodal, and it is most prominent for those planets in multiple systems (blue). The period ratio distribution of these RV multiple systems shows a significant pileup near 2:1 MMR for massive planets. The four vertical lines mark the locations of 3:2, 5:3, 2:1 and 3:1 MMR. }
\label{fig_rvobs}
   \end{center}
\end{figure*}

At the time of writing this paper, there are 409 exoplanets detected with radial velocity (RV) method (exoplanet.org) and about $30\%$ of them reside in multiple planet systems. These RV planets have a wide mass range featured with a bimodal distribution \citep{Pep11} as shown in the left panel of figure\ref{fig_rvobs}. The boundary is at about 0.2 $M_{\rm J}\sim64M_{\oplus}$, which separate the light RV planets (with a media mass of $\sim12M_{\oplus}$) and the massive ones (with a media mass of $\sim1.54M_{\rm J}$). This bimodal distribution may indicate planets undergo different formations and evolutions for the light and massive groups \citep{Mor09}. Interestingly, we find that these two groups may have different period ratio distributions. As shown in the right panels of figure \ref{fig_rvobs}, there is a strong pileup of planet pairs near 2:1 MMR in the massive planet group, which is not seen in the light group.    

\begin{figure*}
\begin{center}
\includegraphics[width=0.5\textwidth,angle=0]{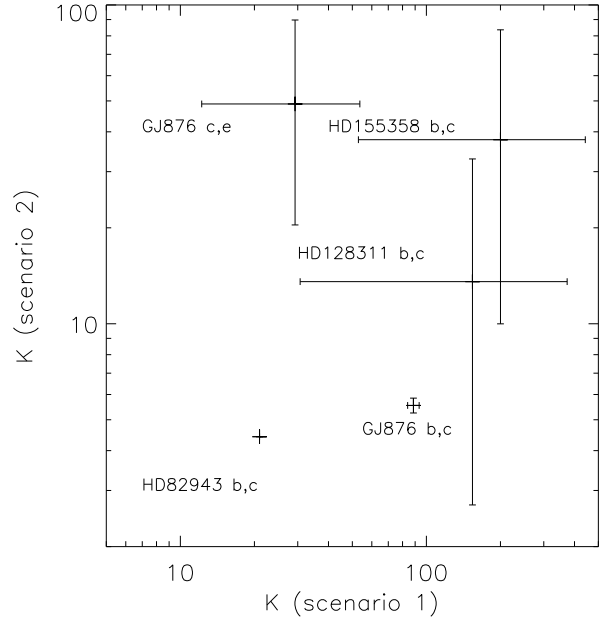}
  \caption{$K-K$ digram for those massive RV planets piled up near 2:1. The horizontal $K$ is the damping ratio $\gamma_{\rm e}/\gamma_{a}$ solved from equation (\ref{deq}) by assuming only the outer planet was subject to orbital damping (scenario 1), i.e., $\gamma_{\rm e2}=K\gamma_{\rm a2}$ and  $\gamma_{\rm e1}=\gamma_{\rm a1}=0$, while the vertical $K$ is the damping ratio solved by assuming both the planets were subject to orbital migration (scenario 2), i.e., $\gamma_{\rm e1}=-K\gamma_{\rm a1}$, $\gamma_{\rm e2}=K\gamma_{\rm a2}$, and $\gamma_{\rm e1}=\gamma_{\rm e2}$, following the two damping scenarios studied in \citet{LP02}. The error bars reflect the uncertainties of their orbital period measurements, except for HD 82943 for which no uncertainty is reported from exoplanet.org. There are another 2 pairs, HD 73526 b and c and 24 Sex b and c, are not plot here because their period uncertainties are too large that could lead to negative $\Delta_{\rm eq}$ in equation (\ref{deq}).  The $K-K$ digram shows a damping ratio ($K$ value) of 1-100 on order of magnitude constrained by the pileup near 2:1 MMR observed in the massive RV sample (see bottom right panel of Fig.\ref{fig_rvobs}). }
\label{fig_k12}
   \end{center}
\end{figure*}

Those massive planets piled up near 2:1 MMR seems unlikely formed in situ within a small annulus, but they are more likely formed with larger distance in a disk then brought into 2:1 MMR through convergent migration. Interestingly, we note that the pile up is just a few percent (in period ratio) wide of the 2:1 MMR center, which is expected from our analytical and numerical prediction with planetary migration (e.g., Fig.\ref{fig_mig}).  
Furthermore, from the location of the pileup (i.e., $\Delta_{\rm eq}$), we can infer the damping ratio between eccentricity and semi major axis during planetary migration (i.e., $K$) by using equation \ref{deq}. 
The result of such an exercise is shown in figure \ref{fig_k12}. Here we considered two migration scenarios. In scenario 1, only the outer planet undergoes migration, i.e., $\gamma_{e2}=K\gamma_{a2}$ and $\gamma_{e1}=\gamma_{a1}=0$. In scenario 2, the inner one migrates outward and the outer one migrates inward, i.e., $\gamma_{e2}=K\gamma_{a2}$, $\gamma_{e1}=-K\gamma_{a1}$ and $\gamma_{a1}=-\gamma_{a2}<0$. As can be seen from figure \ref{fig_k	12}, the $K$ value is constrained in a relative wide range about 1-100 on order of magnitude.  We note this $K$ range is consistent with the hydrodynamical simulations by \citet{Kle04} which predicts a $K$ value of order of unity, and with dynamical modeling of the well-studied system GJ876 by \citet{LP02} which prefers $K=10-100$.

\section{Summary}
\label{sum}
In this paper, we analytically and numerically study the dynamics of planet pairs near first order MMR. Focusing on the evolution of orbital period ratio, we find it could have an asymmetric distribution around the nominal MMR center regardless of whether dissipation is included or not. 

Applying the asymmetric nature of MMR to the \emph{Kepler} planets, we find that, without dissipation, \emph{Kepler} planets' asymmetric period ratio distribution could be \emph{partly} explained for the case of 3:2 MMR but \emph{not} for 2:1 MMR, suggesting that dissipation or other mechanisms may play a more important role in 2:1 than in 3:2 MMR. 

Beside the \emph{Kepler} planets, similar asymmetric feature, i.e., planets piled up wide of MMR, is also seen in RV planets. Nevertheless, planets in multiple RV systems are bimodal distribution on mass, and the pileup is currently only seen in the higher mass group.  The location of the pileup is consistent with the scenario that planetary migration toward MMR, and it infers that the ratio of damping rate between eccentricity and semimajor axis (i.e., K value) during planet migration is $K=1-100$ on order of magnitude for massive planets.

\acknowledgments
JWX thanks the referee for  helpful comments and suggestions, Yanqin Wu and Hanno Rei for valuable discussions  and the \emph{Kepler} team for producing such an invaluable data set. JWX was supported by the National Natural Science Foundation of China
(Nos.\,10833001 and 10925313), PhD training grant of China
(20090091110002), Fundamental Research Funds for the Central
Universities (1112020102) and the Ontario government.
 
%%%%%%%%%%%%%%%%%%%%%%%%%%%%%%%%%%%%%%%%%%%%%%%%%%%%%%%%%%%%%%%%%%%

%%%%%%%%%%%%%%%%%%%%%%%%%%%%%%%%%%%%%%%%%%%%%%%%%%%%%%%%%%%%%%%%%%%
%\newpage
%\bibliography{koi}

\newpage
\appendix
\section{A: Two constants of motion in MMR}
Here we show that the equations \ref{adot1} and/or \ref{const1} are equivalent to the well known constants of motion of MMR. 

For j:j-1 MMR there are two constants of motion in addition to the energy (see chapter 8.8 of \citet{MD99}), i.e.,
\begin{eqnarray}
\Lambda_{1}+(j-1)(\Gamma_{1}+\Gamma_{2})&=&{\rm Const.} \nonumber \\
\Lambda_{2}-j(\Gamma_{1}+\Gamma_{2})&=&{\rm Const.}
\label{const2}
\end{eqnarray}
where, $\Lambda$ and $\Gamma$ are the Poincar\'e momenta (see chapter 2.10 of \citet{MD99}), and the subscript ``1'' and ``2'' denotes the inner and outer planets respectively. Changing the above equation to basic orbital elements, we have
\begin{eqnarray}
m_{1}\sqrt{a_{1}} +(j-1)\left[m_{1}\sqrt{a_{1}}\left(1-\sqrt{1-e_{1}^{2}}\right)+m_{2}\sqrt{a_{2}}\left(1-\sqrt{1-e_{2}^{2}}\right)\right] &=&{\rm Const.} \nonumber \\
m_{2}\sqrt{a_{2}} -j\left[m_{1}\sqrt{a_{1}}\left(1-\sqrt{1-e_{1}^{2}}\right)+m_{2}\sqrt{a_{2}}\left(1-\sqrt{1-e_{2}^{2}}\right)\right] &=&{\rm Const.}
\end{eqnarray}
In the leading term of $e$, we then have
\begin{eqnarray}
m_{1}\sqrt{a_{1}} +(j-1)\left(m_{1}\sqrt{a_{1}}\frac{1}{2}e_{1}^{2}+m_{2}\sqrt{a_{2}}\frac{1}{2}e_{2}^{2}\right) &=&{\rm Const.} \nonumber \\
m_{2}\sqrt{a_{2}} -j\left(m_{1}\sqrt{a_{1}}\frac{1}{2}e_{1}^{2}+m_{2}\sqrt{a_{2}}\frac{1}{2}e_{2}^{2}\right)&=&{\rm Const.}
\end{eqnarray}
Take the differential form of above equations and keep the leading term in $e$, we then have
\begin{eqnarray}
m_{1}\frac{\dot{a_{1}}}{2\sqrt{a_{1}}}+(j-1)\left(m_{1}\sqrt{a_{1}}e_{1}\dot{e_{1}}+m_{2}\sqrt{a_{2}}e_{2}\dot{e_{2}}\right)=0 \nonumber \\
m_{2}\frac{\dot{a_{2}}}{2\sqrt{a_{2}}}-j\left(m_{1}\sqrt{a_{1}}e_{1}\dot{e_{1}}+m_{2}\sqrt{a_{2}}e_{2}\dot{e_{2}}\right)=0
\end{eqnarray}
Using the approximation, $a_{2}/a_{1}\sim [j/(j-1)]^{2/3}$, above equations can be rewritten as equation \ref{adot1}. 
Compared to the original formulas of the constants, the new formulas (Eqn. \ref{adot1}) solve $a_{1}$ and $a_{2}$ out (they are not coupled together as in Eqn.\ref{const2}) and they are dimensionless and simpler.
 
\section{B: Evolution of planetary semi major axes under dissipation near MMR}
Following \citet{Lit12}, it is convenient to introduce the compact eccentricity
\begin{eqnarray}
z_{\rm k} = e_{\rm k}e^{i\varpi_{\rm k}},  
\end{eqnarray}
where $\varpi$ is the longitude of the periastron and $\rm k=1,2$ for the inner and outer planets respectively. In terms of which, the disturbing function can be expressed as
\begin{eqnarray}
R^{j} = \frac{1}{2}\left(f_{1}z_{1}^{*}+f_{2}z_{2}^{*}\right)e^{i\lambda^{j}} +c.c.
\end{eqnarray}
where the superscript `` * '' denotes the complex conjugate of the variable and ``$c.c.$'' denotes the complex conjugate of the proceeding term. Then the eccentricity equation (after adding the damping term) for the two planets is
\begin{eqnarray}
\dot{z_{\rm k}} = -\frac{1}{\sqrt{GM_{\star}}}\frac{2i}{m_{\rm k}\sqrt{a_{\rm k}}}\frac{\partial H}{\partial z_{\rm k}^{*}} - \gamma_{\rm ek}z_{\rm k},
\end{eqnarray}
or specifically,
\begin{eqnarray}
\dot{z_{\rm 1}}= i \rho^{\frac{1}{3}}f_{1} n_{2}\frac{m_{2}}{M_{\star}}e^{i\lambda^{j}}  - \gamma_{\rm e1}z_{\rm 1}, \,\, \, \, \, \, \,
\dot{z_{\rm 2}}= i\rho^{-1}f_{2} n_{1}\frac{m_{1}}{M_{\star}}e^{i\lambda^{j}}  - \gamma_{\rm e2}z_{\rm 2}.
\end{eqnarray}
Adopting the following approximation 
\begin{equation}
\lambda^{j}=j\lambda_{2}-(j-1)\lambda_{1}\sim -j\Delta n_{2}t, \, \, \, \, \gamma_{\rm ek}\ll\Delta n_{2},
\label{approx}
\end{equation}
 then we can solve the eccentricities as 
\begin{eqnarray}
z_{1} =  - \frac{\rho^{1/3}f_{1}}{j\Delta}\frac{m_{2}}{M_{\star}}e^{i\lambda^{j}}\left(1-i\frac{\gamma_{\rm e1}}{j\Delta n_{2}}\right) +z_{\rm free1}, \,\, \, \, \, \, \,
z_{2} =  - \frac{f_{2}}{j\Delta}\frac{m_{1}}{M_{\star}}e^{i\lambda^{j}}\left(1-i\frac{\gamma_{\rm e2}}{j\Delta n_{2}}\right) +z_{\rm free2},
\label{ef}
\end{eqnarray}
where $z_{\rm free1}$ and $z_{\rm free2}$ are free solutions (free eccentricities).

\begin{figure*}
\begin{center}
\includegraphics[width=\textwidth]{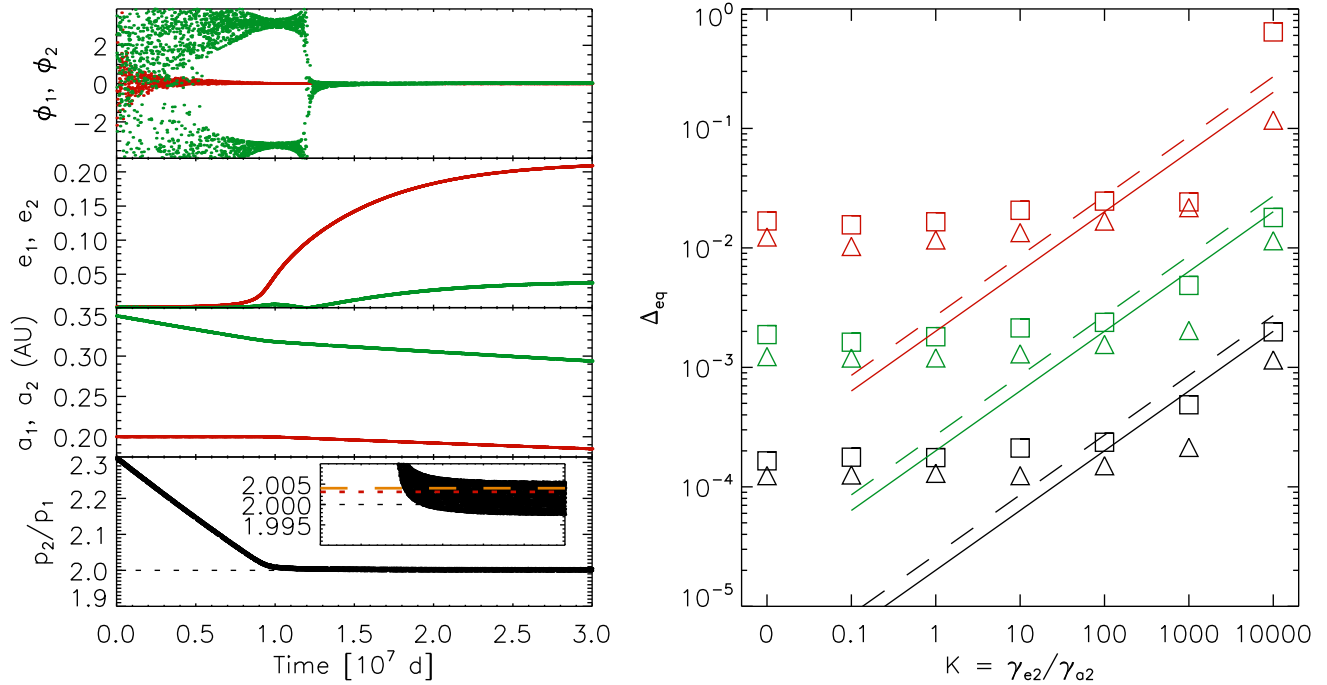}
  \caption{Numerical tests of the asymmetric feature with dissipation. The left four panels show the orbital evolutions (resonance angles, eccentricities, semi major axes and period ratio from top to bottom) of two planets in one simulation with $m_{1}=m_{2}=100M_{\oplus}$ and $\gamma_{e2}=100,\,\gamma_{a2}=10^{-6} \rm d^{-1}$. The outer planet moves inward and then capture in 2:1 MMR with the inner planets. As expected they will finally stay a little bit wider than the MMR center with the mean period ratio equal to  2.003 (red dot line) which is consistent with the estimate from equation \ref{deq}  (orange dashed line in the inserted panel).  The right panels compares the simulated $\Delta_{\rm eq}$ (symbols) to the one predicted from equation \ref{deq} (lines) with different K values, total planetary masses (black, green and red for $ m_{1}+m_{2}=20, 200, 2000 M_{\oplus}$ respectively) and mass ratios (triangle, solid line for $q=1$ and squares and dashed lines for $q=0.25$, respectively). As expected, the analytical results fit roughly well for relative large $K$ ($>10$) and $\Delta_{\rm eq}$ (see also in Appendix B).}
\label{fig_mig}
   \end{center}
\end{figure*}

In terms of the compact eccentricity, the evolution of semi major axes (Eqn.\ref{adot0}) can be rewritten (after adding damping terms) as
\begin{eqnarray}
\frac{\dot{a_{1}}}{a_{1}} = -\frac{(j-1)Gm_{2}}{n_{1}a_{1}^{2}a_{2}} \left[(f_{1}z_{1}^{*}+f_{2}z_{2}^{*})ie^{i\lambda^{j}} + c.c. \right] -\gamma_{a1}
, \,\, \, \, \, \, \,
\frac{\dot{a_{2}}}{a_{2}} = \frac{jGm_{1}}{n_{2}a_{2}^{3}} \left[(f_{1}z_{1}^{*}+f_{2}z_{2}^{*})ie^{i\lambda^{j}} + c.c.  \right] -\gamma_{a1},
\label{adotz}
\end{eqnarray}
which can be finally written as (with the help of Eqn.\ref{ef}), 
\begin{eqnarray}
\frac{\dot{a}_{\rm 1}}{a_{\rm 1}}&=& -\frac{2}{jq\rho^{2/3}}\frac{1}{\Delta^{2}}\left(\frac{m_{2}}{M_{\star}}\right)^{2}\left(q\rho^{\frac{1}{3}}f_{1}^{2}\gamma_{\rm e1}+f_{2}^{2}\gamma_{\rm e2}\right) - \gamma_{\rm a1} + F_{1}
\nonumber \\
\frac{\dot{a}_{\rm 2}}{a_{\rm 2}} &=& \frac{2}{j}\frac{1}{\Delta^{2}}\left(\frac{m_{1}}{M_{\star}}\right)^{2}\left(q\rho^{\frac{1}{3}}f_{1}^{2}\gamma_{\rm e1}+f_{2}^{2}\gamma_{\rm e2}\right) - \gamma_{\rm a2}+ F_{2},
\label{da5}
\end{eqnarray}
where $F_{1}$ and $F_{2}$ are the terms caused by the free eccentricities, i.e., 
\begin{eqnarray}
F_{1} = -\frac{(j-1)Gm_{2}}{n_{1}a_{1}^{2}a_{2}} (Z_{\rm free}^{*}ie^{i\lambda^{j}} + c.c.)
, \,\, \, \, \, \, \,
F_{2} = \frac{jGm_{1}}{n_{2}a_{2}^{3}} (Z_{\rm free}^{*}ie^{i\lambda^{j}} + c.c.),
\label{adotz}
\end{eqnarray}
for $Z_{\rm free}=f_{1}z_{1}+f_{2}z_{2}$ defined as the free eccentricities of the system.
. In the case where it is not too close to MMR (modest $\Delta$) and eccentricity damping is efficient (large $\gamma_{e_{\rm k}}$), $Z_{\rm free}\sim0$ and thus the two oscillation terms $F_{1}$ and $F_{2}$ can be ignored (i.e., Eqn.\ref{da4} and \ref{deq}). Otherwise, if it is very close to MMR (very small $\Delta$) and the eccentricity damping is weak, then the system could get significant free eccentricities (probably by approaching the separatrix), thus $F_{1}$ and $F_{2}$ cannot be ignored and the equilibrium cannot be well estimated by using equation \ref{deq} (see also in Fig.\ref{fig_mig}).

%%%%%%%%%%%%%  Figure 1 %%%%%%%%%%%%%%%%
\newpage
%%%%%%%%%%%%%%%%%%%%%%%%%%%%%%%%%%

%%%%%%%%%%%%%  Figure 2 %%%%%%%%%%%%%%%%
%%%%%%%%%%%%%%%%%%%%%%%%%%%%%%%%%%

%%%%%%%%%%%%%  Figure 3 %%%%%%%%%%%%%%%%

%%%%%%%%%%%%%%%%%%%%%%%%%%%%%%%%%%

%%%%%%%%%%%%%  Figure 4 %%%%%%%%%%%%%%%%

%%%%%%%%%%%%%%%%%%%%%%%%%%%%%%%%%%

%%%%%%%%%%%%%  Figure 5 %%%%%%%%%%%%%%%%

%%%%%%%%%%%%%%%%%%%%%%%%%%%%%%%%%%

%%%%%%%%%%%%%  Figure 6 %%%%%%%%%%%%%%%%

%%%%%%%%%%%%%  Figure 7 %%%%%%%%%%%%%%%%

%%%%%%%%%%%%%  Figure 8 %%%%%%%%%%%%%%%%

%\end{CJK*}
\end{document}